\documentclass[12pt,a4paper]{article}
\usepackage{amsmath,amssymb}
\usepackage{url,cite,slashed}
\usepackage[dvipdfmx]{graphicx}
\usepackage{subfigure}
\usepackage{color}

\begin{document}

\renewcommand{\thefootnote}{\fnsymbol{footnote}} % For titlepage
\begin{titlepage}

\begin{center}

\hfill UT--16--08\\

\vskip .75in

{\Large \bf 
ATLAS on-$Z$ Excess Through Vector-Like Quarks
}

\vskip .75in

{\large
Motoi Endo
and 
Yoshitaro Takaesu
}

\vskip 0.25in

{\em Department of Physics, University of Tokyo, Tokyo 113--0033, Japan}

\end{center}

\vskip .5in

\begin{abstract}
\noindent We investigate the possibility that the excess observed in the leptonic-$Z +{\rm
 jets} +\slashed{E}_T$ ATLAS SUSY search is due to pair productions of a vector-like 
quark $U$ decaying to the first-generation quarks and $Z$ boson. We find that the excess can be explained within the 2$\sigma$ (up to 1.4$\sigma$) level while evading the constraints from the other LHC searches. The preferred range
 of the mass and branching ratio are $610 < m_{U} < 760$\,GeV and ${\rm Br}(U \rightarrow Z q) > 0.3$--$0.45$, respectively.
\end{abstract}

\end{titlepage}

\setcounter{page}{1}
\renewcommand{\thefootnote}{\#\arabic{footnote}}
\setcounter{footnote}{0}

%%%%%%%%%%%%%%%%%%%%%%%%%%%%%%%%%%%%%%%%%%%%%%%%%%%%
%\section{Introduction}
\noindent{\itshape\bfseries  Introduction}\indent
After the LHC 8\,TeV run, an excess has been reported in the leptonic-$Z +{\rm jets} +\slashed{E}_T$ (``on-$Z$'') channel by the ATLAS collaboration~\cite{Aad:2015wqa}. 
The observed number of the signal events are 16 and 13 for the final-state electrons and muons, respectively, whereas the standard model (SM) predicts $4.2 \pm 1.6$ and $6.4 \pm 2.2$. 
The discrepancy corresponds to the $3\sigma$ level, which stimulates many
theoretical
 studies~\cite{Barenboim:2015afa,Vignaroli:2015ama,Ellwanger:2015hva,Allanach:2015xga,Kobakhidze:2015dra,Cao:2015ara,Cahill-Rowley:2015cha,Lu:2015wwa,Liew:2015hsa,Cao:2015zya,Collins:2015boa,Harigaya:2015pma,Lu:2016xwv}.\footnote{
Although no excess has been observed by the CMS collaboration~\cite{Khachatryan:2015lwa,CMS:2015bsf}, the ATLAS collaboration has reported a new result based on the 13\,TeV data recently~\cite{ATLAS13TeV}, which shows the deviation at the 2.2$\sigma$ level. 
} 

The on-$Z$ signal was investigated originally to search for the supersymmetry (SUSY)~\cite{Aad:2015wqa}, and most of the theoretical works have been performed within the framework of SUSY.
In this letter, we instead consider models with vector-like (VL) quarks as an alternative scenario.
The VL particles are predicted in new physics models, e.g., the little Higgs models~\cite{ArkaniHamed:2001nc,ArkaniHamed:2002pa,ArkaniHamed:2002qx,Low:2002ws,Kaplan:2003uc} and the composite Higgs models~\cite{Kaplan:1983fs,Kaplan:1983sm,Kaplan:1991dc,Agashe:2004rs,Contino:2006qr,Contino:2006nn,Carena:2007ua}.
We assume that the VL quarks are pair-produced directly at the LHC by the QCD interactions.\footnote{
VL quark productions through a heavy-gluon decay are studied in Ref.~\cite{Vignaroli:2015ama,Lu:2016xwv}.
}
Then, they decay into SM quarks and bosons through their mixings with the SM quarks, since otherwise they become stable and conflict with the cosmology and experiments~\cite{Smith:1979rz,Smith:1982qu,Hemmick:1989ns,Starkman:1990nj,Verkerk:1991jf,Yamagata:1993jq,Kudo:2001ie,Mack:2007xj,Mack:2012ju}.
The decay modes involve productions of the on-shell $Z$ bosons, which contribute to the ATLAS signal.
Since the branching ratios of the VL quark depend on details of the models, they are supposed to be free parameters in this letter, and we examine whether this scenario works as a candidate of the on-$Z$ excess.

The models with the VL quarks may be distinguished from the SUSY ones if
signal event distributions are precisely measured.
In particular, the SUSY models tend to predict events with larger jet
multiplicity, e.g., through the gluino pair production, $pp \to \tilde g\tilde g, \tilde g \to q\bar{q}\tilde\chi^0_1 \to q\bar{q} Z\tilde{G}$~\cite{Aad:2015wqa,Barenboim:2015afa,Allanach:2015xga}, where $\tilde\chi^0_1$ is the lightest neutralino, and
$\tilde{G}$ is the gravitino.\footnote{
Squark productions can predict lower jet multiplicity as $pp \to \tilde q\tilde q^*, \tilde q \to q\tilde\chi^0_i \to q Z\tilde\chi^0_1$~\cite{Cahill-Rowley:2015cha,Cao:2015zya}, where $\tilde\chi^0_i$ is a heavier neutralino.
}
In contrast, the VL quark $U$ decays into less-multiple jets through, e.g., $U \to qZ$.
Although the current integrated luminosity at the LHC is not large enough to determine the distributions, the data may prefer a lower jet multiplicity.
Thus, we also study the event distributions in the VL quark models.
\\

%%%%%%%%%%%%%%%%%%%%%%%%%%%%%%%%%%%%%%%%%%%%%%%%%%%%
\noindent{\itshape\bfseries  Model}\indent
We extend the SM by introducing a VL quark which has 
the electric charge of $2/3$ and only decays to the first-generation quarks. Discussions for VL quarks carrying the electric
charge of $-1/3$ or decaying also to the second-generation quarks go along
the same lines.\footnote{On the other hand, decays to the third-generation quarks are severely constrained by the LHC Run-I searches. For example, VL quarks of the mass less than 800\,GeV have been already excluded~\cite{Aad:2014efa,Aad:2015gdg,Aad:2015kqa}.}

The interactions of the VL quark with gluons and photons are
governed by the gauge symmetries.  
On the other hand,
 the interactions to the weak gauge and Higgs bosons are model
dependent, and we employ an effective-model approach. The interaction of the VL quark with the weak gauge and Higgs bosons are parameterized as~\cite{Buchkremer:2013bha}
\begin{align}
   {\cal L}_{\rm eff} =& \eta \left(
 \kappa_W \frac{g}{\sqrt{2}}\bar{U}_L
 W^+_\mu \gamma^\mu d_L 
+\kappa_Z \frac{g}{2 c_W}\bar{U}_L Z_\mu
 \gamma^\mu u_L  -\kappa_h \frac{M_U}{v}\bar{U}_R \,h \,u_L \right)
+ {\rm h.c.},
\label{eq:lagrangian}
\end{align}
where $g$ is the $SU(2)_L$ gauge coupling constant, $c_W$ cosine of the
Weinberg angle, $v \simeq 246$\,GeV the vacuum expectation value of the Higgs field,
$M_U$ the mass of the VL quark, and
\begin{align}
  \eta \equiv& \sqrt{16\pi \frac{v^2}{M_U^3}\Gamma_U}
\end{align}
with $\Gamma_U$ being the total width of
the VL quark. 
The VL couplings with the first-generation quarks are constrained to be less than ${\cal O}(0.01)$~\cite{Fajfer:2013wca}.
In order to avoid this constraint, $\eta$ is taken to be small. The following discussion does not depend on its detail as long as the VL quark decays promptly. 
For simplicity, we only consider the case that the VL quark couples to the left-handed light quarks.\footnote{
The following study does not depend on this assumption. 
The chirality structure may be identified by investigating
angular correlations of the final-state particles.}
In the following discussion, we take the branching ratio of the VL
quark, ${\rm Br}(U\to V q)$, as free parameters by expressing
the $\kappa_V$ ($V = W, Z, h$) as
\begin{align}
 \kappa_V =& \sqrt{\frac{{\rm Br}(U \rightarrow V q)}{\gamma_V}},
\end{align}
where
 $q = u, d$ and
\begin{subequations}
\begin{align}
\gamma_W =& \left(1 -\frac{m_W^2}{M_U^2}\right)^2 \left(1
 +\frac{m_W^2}{M_U^2} -2\frac{m_W^4}{M_U^4} \right) +{\cal
 O}\left( \frac{m_d^2}{M_U^2} \right), \\
\gamma_Z =& \frac{1}{2} \left( 1-\frac{m_Z^2}{M_U^2}  \right)^2
\left(1 +\frac{m_Z^2}{M_U^2} -2\frac{m_Z^4}{M_U^4} \right) +{\cal
 O}\left( \frac{m_u^2}{M_U^2} \right), \\
\gamma_h =& \frac{1}{2} \left( 1 -\frac{m_h^2}{M_U^2} \right)^2 +{\cal
 O}\left( \frac{m_u^2}{M_U^2} \right).
\end{align}
\end{subequations}
Note that the branching ratios are independent of $\eta$.
\\

\noindent{\itshape\bfseries  Analysis} \indent
We consider pair-production processes of the VL quark $U$, decaying to the first-generation quarks along with $Z, W$
or Higgs bosons at the 8\,TeV LHC:
\begin{equation}
 p p \rightarrow U \bar{U}, \hspace{1em}  U \rightarrow Z
  u, W^+ d, h u.
\end{equation}
Those processes are generated at the tree level using MadGraph5\_aMC@NLO
v2.3~\cite{Alwall:2014hca}. The model file of the VL quark~\cite{Buchkremer:2013bha,VLQmodel} is implemented via FeynRules v2.3~\cite{Alloul:2013bka}. 
The generated events are passed to PYTHIA v6.428\cite{Sjostrand:2006za} for decaying the $Z, W$ and Higgs bosons
as well as showering and hadronization, and then interfaced to the
Delphes3-based detector
simulator in CheckMATE v1.2.1~\cite{deFavereau:2013fsa,Drees:2013wra}, which is
tuned to reproduce the performance of the ATLAS detector.
The cross sections of the VL-quark pair productions are
estimated at the next-to-next-to-leading order (NNLO) accuracy with
Hathor v2.0~\cite{Aliev:2010zk}. MSTW 2008 NNLO (68\%CL) PDF~\cite{Martin:2009iq} is used with the factorization and renormalization scales set at the
mass of the VL quark.

We then analyze the generated events following the LHC analyses
below
%\footnote{We use those analyses implemented in
%CheckMATE~\cite{Drees:2013wra,Cao:2015ara}.}:
~\cite{Drees:2013wra,Cao:2015ara}.
\begin{itemize}
 \item ATLAS search for leptonic-$Z +{\rm jets} +\slashed{E}_T$ signal~\cite{Aad:2015wqa}
 \item ATLAS search for 2-6 {\rm jets} $+\slashed{E}_T$ signal~\cite{Aad:2014wea}
 \item CMS search for leptonic-$Z +{\rm jets} +\slashed{E}_T$ signal
       (on-$Z$ signal region)~\cite{Khachatryan:2015lwa}.
\end{itemize}
The first ATLAS analysis is used to search for parameter regions
where the ATLAS excess is reproduced, and
the other analyses are used to constrain the parameter space.
We also checked that the other LHC searches implemented in CheckMATE v1.2.1 do not give severer constraints.
\\

\begin{figure}[t]
\centering
\includegraphics[scale=0.8]{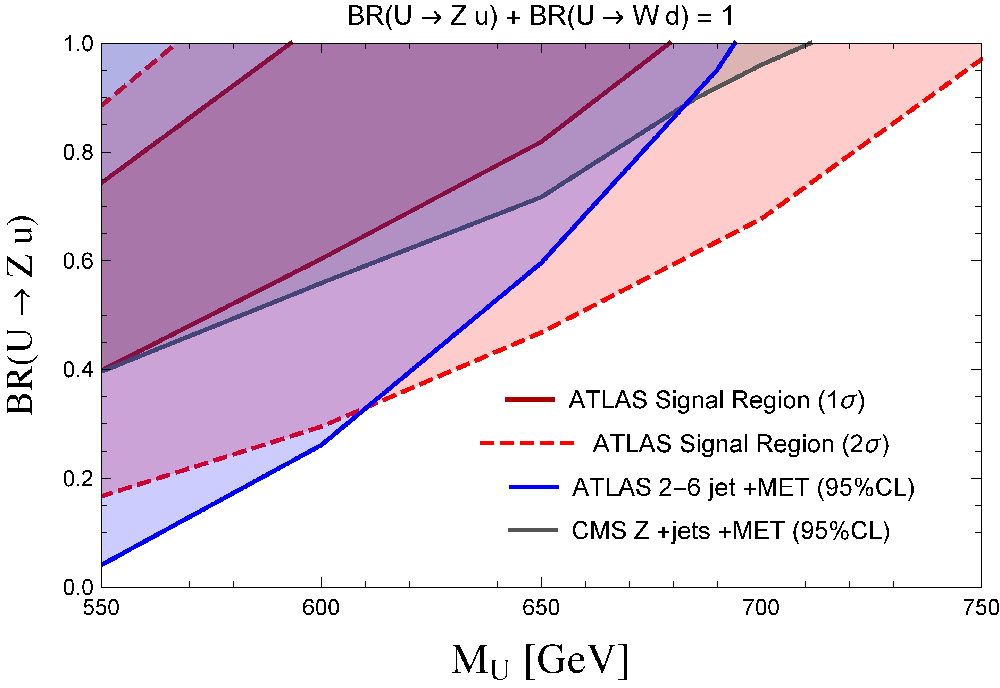} 
 \includegraphics[scale=0.8]{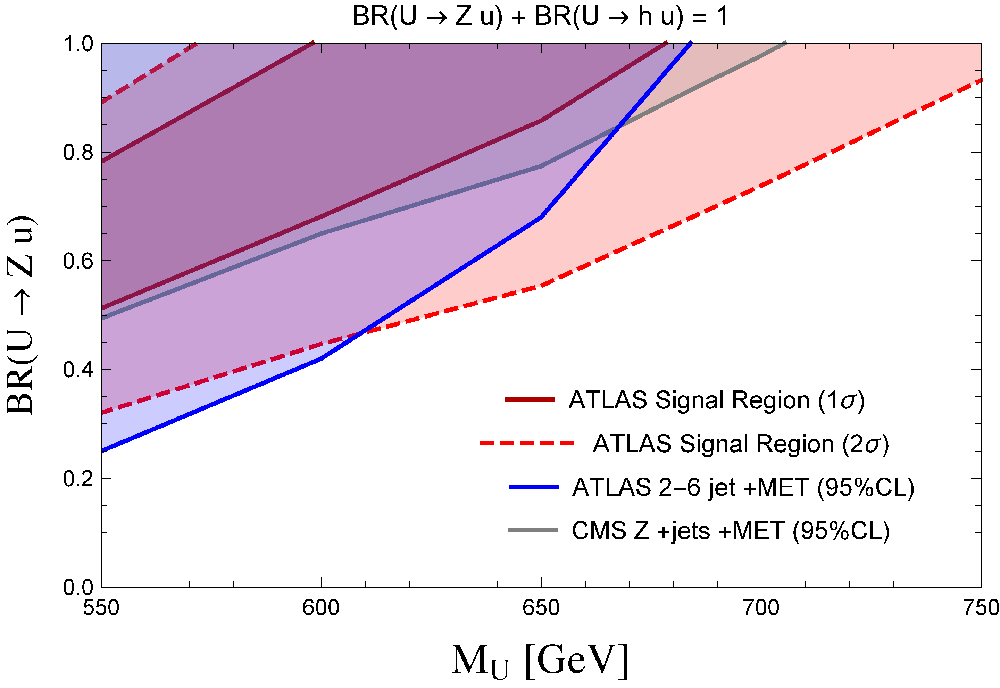}
 \caption{The ATLAS signal region and LHC constraints in
 the ${\rm Br}(U \rightarrow Zu$) vs.~$M_{U}$ plane. 
In the dark-red (light-red) region, the ATLAS on-$Z$ excess~\cite{Aad:2015wqa} is explained within the $1 \sigma$ (2$\sigma$) level, while the gray- and
 blue-shaded regions are excluded at 95\% C.L.~by the CMS leptonic-$Z \,+$jets $+\slashed{E}_T$ search~\cite{Khachatryan:2015lwa} and ATLAS 2-6
 jet $+\slashed{E}_T$ search~\cite{Aad:2014wea}, respectively. The
 vector-like quark is assumed to decay with $Z$ or $W$ boson emission
 (left panel) and $Z$ or Higgs boson emission (right panel).}
\label{fig:limit}
\end{figure}

%%%%%%%%%%%%%%%%%%%%%%%%%%%%%%%%%%%%%%%%%%%%%%%%%%%%
\noindent{\itshape\bfseries Results}\indent
In Fig.~\ref{fig:limit} we show the parameter regions where the
ATLAS excess in the
leptonic-$Z +{\rm jets} +\slashed{E}_T$ channel\cite{Aad:2015wqa} is reproduced within
1$\sigma$ (dark-red shaded region) and 2$\sigma$ (light-red shaded region), corresponding to the signal
event number of $12.1 \leq N_{\rm sig} \leq 24.7$ and $5.8 \leq N_{\rm sig} \leq 31$~\cite{Cao:2015ara}, respectively.
The model parameter space is spanned by the mass of the VL
quark and the branching ratio Br$(U
\rightarrow Z u)$.
In the figure we assume that the VL quark decays
via $Z$ or $W$ boson emission ($Z$-$W$ decay in the left panel) or via $Z$ or Higgs boson emission
($Z$-Higgs decay in the right panel).
The excluded regions from
the other LHC analyses, i.e., the CMS search in the leptonic-$Z +{\rm jets}
+\slashed{E}_T$ final states~\cite{Khachatryan:2015lwa} and ATLAS search in the 2-6 jets
+$\slashed{E}_T$ final states~\cite{Aad:2014wea} are also shown as the gray- and blue-shaded
regions, respectively. 
We see that the VL quark model with a mass of $610 \lesssim
M_U \lesssim 760$\,GeV can explain the ATLAS excess within $2\sigma$ (up to 1.4$\sigma$) in
both the $Z$-$W$ and $Z$-Higgs decay cases. 

The CMS search in the leptonic-$Z \,+$jets $+\slashed{E}_T$ final states
excludes the parameter space of $m_{U} \lesssim 710$\,GeV and ${\rm Br}(U \rightarrow Z u) \gtrsim 0.4$--$0.5$.
Since the same final states as the ATLAS on-$Z$ excess are investigated, the 95\% C.L.~exclusion lines are roughly parallel to the 1$\sigma$ and 2$\sigma$ ATLAS signal-region contours.
The CMS exclusion region covers the whole 1$\sigma$ ATLAS signal region (dark-red shaded
 region) for $M_U > 550$\,GeV, 
 but still allowing the 2$\sigma$ signal region (light-red shaded
 region) for wide range of
 the VL quark mass due to the large uncertainty on the signal event
 number of the ATLAS
 on-$Z$ excess. 

On the other hand, the 2-6
jets $+\slashed{E}_T$ search is sensitive to a smaller $Z$-branching region
 and excludes the whole 1$\sigma$ ATLAS signal
region as well as a part of the
2$\sigma$ signal region for $M_U \lesssim 680$\,GeV. In the $Z$-$W$ decay
case, the model parameter space of $M_U < 550$\,GeV is almost excluded, while there remains an allowed region for ${\rm Br}(U \rightarrow Z u) \lesssim 0.2$ in the $Z$-Higgs decay case. 
This is because the 2-6 jets $+\slashed{E}_T$ search is not so sensitive
to the parameter space dominated by
the Higgs-involving $U$ decays, which basically do not leave large
$\slashed{E}_T$. 
The CMS constraint is stronger than the ATLAS 2-6 jets
$+\slashed{E}_T$ search for Br($U \rightarrow Z u$) $ \gtrsim 0.8$.

\begin{figure}[t]
\centering
\includegraphics[scale=0.8]{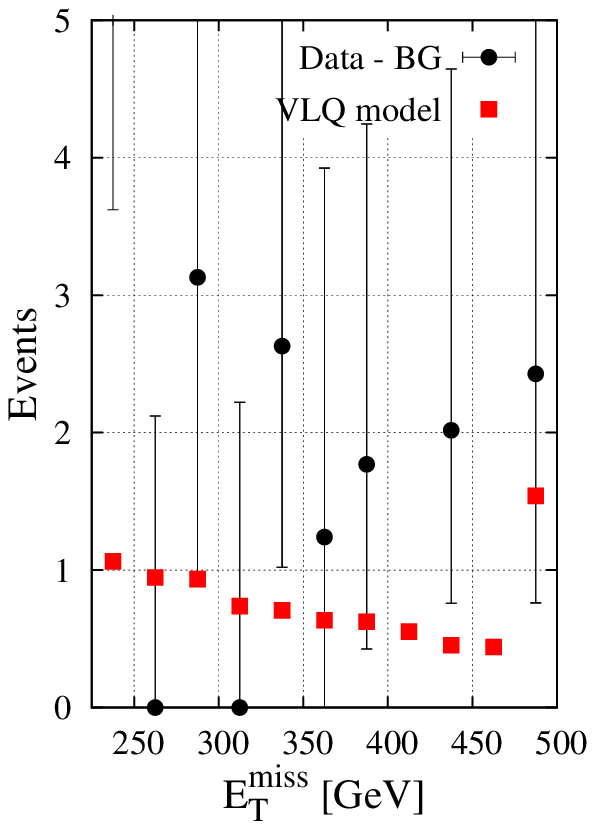} 
\includegraphics[scale=0.8]{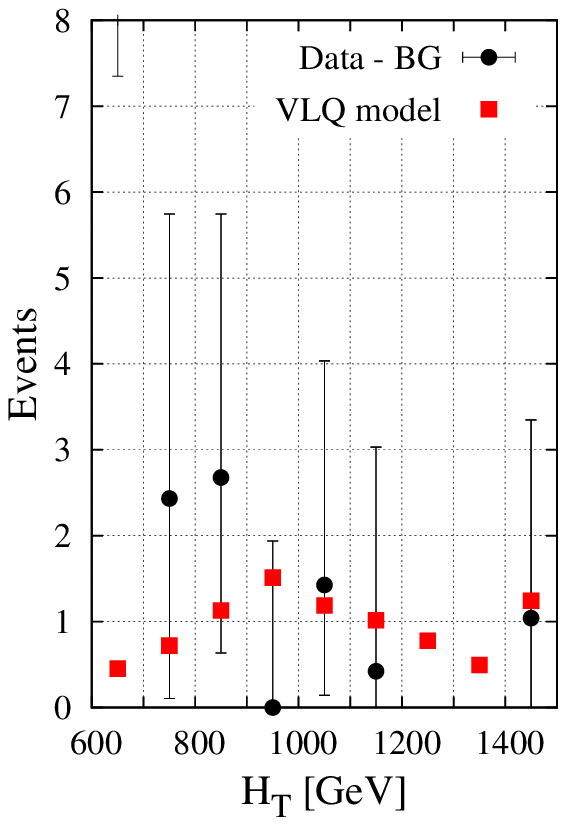}
\includegraphics[scale=0.8]{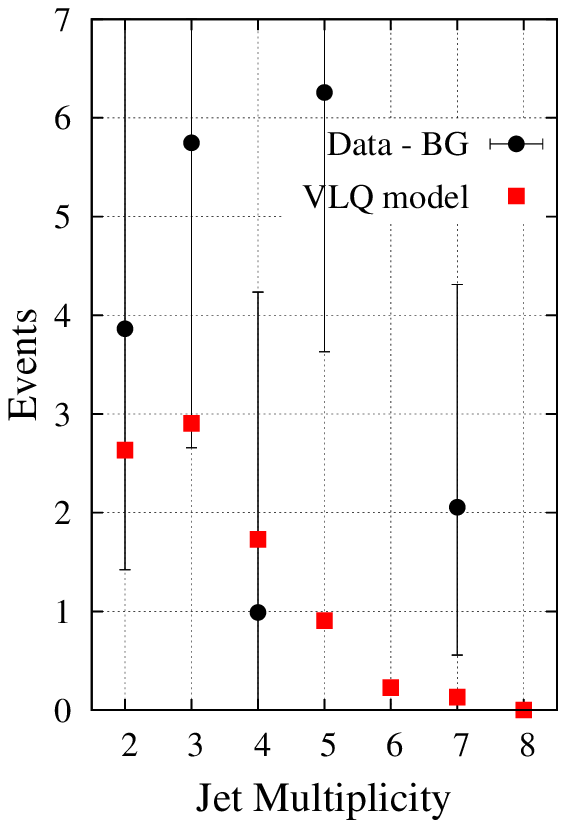}  
 \caption{The distributions of the $\slashed{E}_T, H_T$ and jet multiplicity
 predicted by the VL quark model with $M_U = 680$\,GeV and ${\rm Br}(U
 \rightarrow Z u) = 0.8$ (red boxes). The experimental data of the ATLAS on-$Z$
 excess with the expected SM backgrounds being subtracted~\cite{Cao:2015zya} are also shown
 (black dots with error bars). The highest bins contain overflow events.}
\label{fig:dist}
\end{figure}

Next we show the $\slashed{E}_T, H_T$ and jet-multiplicity distributions predicted by the VL quark model with $M_U = 680$\,GeV and ${\rm Br}(U
  \rightarrow Z u) = 0.8$ for the $Z$-$W$ decay case (red boxes) in Fig.~\ref{fig:dist}. 
This model point gives 8.3 signal events, which agrees with the ATLAS on-$Z$ excess at 1.6$\sigma$. 
The black dots show the ATLAS data with the expected SM backgrounds being subtracted (taken from Ref.~\cite{Cao:2015zya}).
In the figure, all the model distributions show marginal agreements with the ATLAS ones.
For the $\slashed{E}_T$, $H_T$ and jet-multiplicity distributions, $\chi^2/{\rm d.o.f.} = 7.7/9$, 5.4/7 and 7.0/5, respectively.
It is mentioned that the jet-multiplicity distribution of the VL quark model peaks around 2-3
number of jets, which may be a distinguishable feature of the model. 
Future LHC Run-II data is expected to reveal the detailed nature of the excess.
\\

%%%%%%%%%%%%%%%%%%%%%%%%%%%%%%%%%%%%%%%%%%%%%%%%%%%%
\noindent{\itshape\bfseries  Conclusion}\indent
In this letter we have investigated the possibility that the excess
observed in the ATLAS SUSY search in the leptonic-$Z +{\rm
 jets} +\slashed{E}_T$ final states is due to pair productions of the VL 
quark $U$, which only decays to the first-generation quarks. 
We find that the excess can be explained
within the 2$\sigma$ (up to 1.4$\sigma$) level while evading the constraints from the other LHC searches such as the CMS leptonic-$Z +{\rm jets} +\slashed{E}_T$ and ATLAS 2-6 jets $+\slashed{E}_T$ searches. The $2\sigma$ preferred range of the VL-quark mass and branching ratio of the $Z$-boson involving decay are $610 \lesssim M_{U} \lesssim 760$\,GeV and ${\rm Br}(U \rightarrow Z u) \gtrsim 0.3$--$4.5$ (depending on the decay modes of the VL quark), respectively. The $\slashed{E}_T, H_T$ and jet-multiplicity
 distributions predicted by the VL-quark model show marginal agreements
 with those of the ATLAS excess. 
%Due to the large experimental uncertainties, it is still early to state anything from the distributions.
In conclusion, there is room for VL quark models to explain the ATLAS on-$Z$ excess, and upcoming results from the 13\,TeV LHC would confirm or refute the VL quark interpretation of the excess.
\\

%%%%%%%%%%%%%%%%%%%%%%%%%%%%%%%%%%%%%%%%%%%%%%%%%%%%
\noindent{\itshape\bfseries  Acknowledgments}
This work was supported by JSPS KAKENHI Grant No.~25105011 (M.E.) and
No.~23104008  (Y.T.).

%%%%%%%%%%%%%%%%%%%%%%%%%%%%%%%%%%%%%%%%%%%%%%%%%%%%
\providecommand{\href}[2]{#2}
\begingroup\raggedright

\endgroup

\begin{thebibliography}{99}
\bibitem{Aad:2015wqa} 
  G.~Aad {\it et al.} [ATLAS Collaboration],
  %``Search for supersymmetry in events containing a same-flavour
	%opposite-sign dilepton pair, jets, and large missing transverse
	%momentum in $\sqrt{s}=8$  TeV pp collisions with the ATLAS
	%detector,''
  Eur.\ Phys.\ J.\ C {\bf 75}, no. 7, 318 (2015)
  [Eur.\ Phys.\ J.\ C {\bf 75}, no. 10, 463 (2015)]
  doi:10.1140/epjc/s10052-015-3661-9, 10.1140/epjc/s10052-015-3518-2
  [arXiv:1503.03290 [hep-ex]].

%\cite{Barenboim:2015afa}
\bibitem{Barenboim:2015afa} 
  G.~Barenboim, J.~Bernabeu, V.~A.~Mitsou, E.~Romero and O.~Vives,
  %``METing SUSY on the Z peak,''
  Eur.\ Phys.\ J.\ C {\bf 76}, 57 (2016)
  doi:10.1140/epjc/s10052-016-3901-7
  [arXiv:1503.04184 [hep-ph]].

%\cite{Vignaroli:2015ama}
\bibitem{Vignaroli:2015ama} 
  N.~Vignaroli,
  %``$Z$-peaked excess from heavy gluon decays to vectorlike quarks,''
  Phys.\ Rev.\ D {\bf 91}, no. 11, 115009 (2015)
  doi:10.1103/PhysRevD.91.115009
 [arXiv:1504.01768 [hep-ph]].

%\cite{Ellwanger:2015hva}
\bibitem{Ellwanger:2015hva} 
  U.~Ellwanger,
  %``Possible explanation of excess events in the search for jets,
	%missing transverse momentum and a Z boson in pp collisions,''
  Eur.\ Phys.\ J.\ C {\bf 75}, no. 8, 367 (2015)
  doi:10.1140/epjc/s10052-015-3591-6
  [arXiv:1504.02244 [hep-ph]].

%\cite{Allanach:2015xga}
\bibitem{Allanach:2015xga} 
  B.~Allanach, A.~Raklev and A.~Kvellestad,
  %``Consistency of the recent ATLAS $Z+E_T^{\rm miss}$ excess in a
	%simplified GGM model,''
  Phys.\ Rev.\ D {\bf 91}, 095016 (2015)
  doi:10.1103/PhysRevD.91.095016
  [arXiv:1504.02752 [hep-ph]].

%\cite{Kobakhidze:2015dra}
\bibitem{Kobakhidze:2015dra} 
  A.~Kobakhidze, N.~Liu, L.~Wu and J.~M.~Yang,
  %``ATLAS Z-peaked excess in the MSSM with a light sbottom or stop,''
  Phys.\ Rev.\ D {\bf 92}, no. 7, 075008 (2015)
  doi:10.1103/PhysRevD.92.075008
  [arXiv:1504.04390 [hep-ph]].

%\cite{Cao:2015ara}
\bibitem{Cao:2015ara} 
  J.~Cao, L.~Shang, J.~M.~Yang and Y.~Zhang,
  %``Explanation of the ATLAS Z-Peaked Excess in the NMSSM,''
  JHEP {\bf 1506}, 152 (2015)
  doi:10.1007/JHEP06(2015)152
  [arXiv:1504.07869 [hep-ph]].

%\cite{Cahill-Rowley:2015cha}
\bibitem{Cahill-Rowley:2015cha} 
  M.~Cahill-Rowley, J.~L.~Hewett, A.~Ismail and T.~G.~Rizzo,
  %``ATLAS Z\UTF{2009}+ missing transverse energy excess in the MSSM,''
  Phys.\ Rev.\ D {\bf 92}, 075029 (2015)
  doi:10.1103/PhysRevD.92.075029
  [arXiv:1506.05799 [hep-ph]].

%\cite{Lu:2015wwa}
\bibitem{Lu:2015wwa} 
  X.~Lu, S.~Shirai and T.~Terada,
  %``ATLAS $Z$ Excess in Minimal Supersymmetric Standard Model,''
  JHEP {\bf 1509}, 204 (2015)
  doi:10.1007/JHEP09(2015)204
  [arXiv:1506.07161 [hep-ph]].

%\cite{Liew:2015hsa}
\bibitem{Liew:2015hsa} 
  S.~P.~Liew, A.~Mariotti, K.~Mawatari, K.~Sakurai and M.~Vereecken,
  %``Z-peaked excess in goldstini scenarios,''
  Phys.\ Lett.\ B {\bf 750}, 539 (2015)
  doi:10.1016/j.physletb.2015.09.035
  [arXiv:1506.08803 [hep-ph]].

%\cite{Cao:2015zya}
\bibitem{Cao:2015zya} 
  J.~Cao, L.~Shang, J.~M.~Yang and Y.~Zhang,
  %``Explanation of the ATLAS Z-peaked excess by squark pair production
	%in the NMSSM,''
  JHEP {\bf 1510}, 178 (2015)
  doi:10.1007/JHEP10(2015)178
  [arXiv:1507.08471 [hep-ph]].

%\cite{Collins:2015boa}
\bibitem{Collins:2015boa} 
  J.~H.~Collins, J.~A.~Dror and M.~Farina,
  %``Mixed Stops and the ATLAS on-Z Excess,''
  Phys.\ Rev.\ D {\bf 92}, no. 9, 095022 (2015)
  doi:10.1103/PhysRevD.92.095022
  [arXiv:1508.02419 [hep-ph]].

%\cite{Harigaya:2015pma}
\bibitem{Harigaya:2015pma} 
  K.~Harigaya, M.~Ibe and T.~Kitahara,
  %``ATLAS on-Z Excess via gluino-Higgsino-singlino decay chains in the
	%NMSSM,''
  JHEP {\bf 1601}, 030 (2016)
  doi:10.1007/JHEP01(2016)030
  [arXiv:1510.07691 [hep-ph]].

%\cite{Lu:2016xwv}
\bibitem{Lu:2016xwv} 
  X.~Lu, S.~Shirai and T.~Terada,
  %``Testing ATLAS Z+MET Excess with LHC Run 2,''
  arXiv:1601.05777 [hep-ph].

\bibitem{Khachatryan:2015lwa} 
  V.~Khachatryan {\it et al.} [CMS Collaboration],
  %``Search for Physics Beyond the Standard Model in Events with Two
	%Leptons, Jets, and Missing Transverse Momentum in pp Collisions
	%at sqrt(s) = 8 TeV,''
  JHEP {\bf 1504}, 124 (2015)
  doi:10.1007/JHEP04(2015)124
  [arXiv:1502.06031 [hep-ex]].

%\cite{CMS:2015bsf}
\bibitem{CMS:2015bsf}
  CMS Collaboration [CMS Collaboration],
  %``Search for new  physics in final states with two opposite-sign
	%same-flavor leptons, jets and missing transverse momentum in pp
	%collisions at sqrt(s)=13 TeV,''
  CMS-PAS-SUS-15-011.

\bibitem{ATLAS13TeV} 
  The ATLAS collaboration,
  %``A search for Supersymmetry in events containing a leptonically
	%decaying $Z$ boson, jets and missing transverse momentum in
	%$\sqrt{s}=13~$TeV $pp$ collisions with the ATLAS detector,''
  ATLAS-CONF-2015-082.

%\cite{ArkaniHamed:2001nc}
\bibitem{ArkaniHamed:2001nc} 
  N.~Arkani-Hamed, A.~G.~Cohen and H.~Georgi,
  %``Electroweak symmetry breaking from dimensional deconstruction,''
  Phys.\ Lett.\ B {\bf 513}, 232 (2001)
  doi:10.1016/S0370-2693(01)00741-9
  [hep-ph/0105239].

%\cite{ArkaniHamed:2002pa}
\bibitem{ArkaniHamed:2002pa} 
  N.~Arkani-Hamed, A.~G.~Cohen, T.~Gregoire and J.~G.~Wacker,
  %``Phenomenology of electroweak symmetry breaking from theory space,''
  JHEP {\bf 0208}, 020 (2002)
  [hep-ph/0202089].

%\cite{ArkaniHamed:2002qx}
\bibitem{ArkaniHamed:2002qx} 
  N.~Arkani-Hamed, A.~G.~Cohen, E.~Katz, A.~E.~Nelson, T.~Gregoire and J.~G.~Wacker,
  %``The Minimal moose for a little Higgs,''
  JHEP {\bf 0208}, 021 (2002)
  doi:10.1088/1126-6708/2002/08/021
  [hep-ph/0206020].

%\cite{Low:2002ws}
\bibitem{Low:2002ws} 
  I.~Low, W.~Skiba and D.~Tucker-Smith,
  %``Little Higgses from an antisymmetric condensate,''
  Phys.\ Rev.\ D {\bf 66}, 072001 (2002)
  doi:10.1103/PhysRevD.66.072001
  [hep-ph/0207243].

%\cite{Kaplan:2003uc}
\bibitem{Kaplan:2003uc} 
  D.~E.~Kaplan and M.~Schmaltz,
  %``The Little Higgs from a simple group,''
  JHEP {\bf 0310}, 039 (2003)
  doi:10.1088/1126-6708/2003/10/039
  [hep-ph/0302049].

%\cite{Kaplan:1983fs}
\bibitem{Kaplan:1983fs} 
  D.~B.~Kaplan and H.~Georgi,
  %``SU(2) x U(1) Breaking by Vacuum Misalignment,''
  Phys.\ Lett.\ B {\bf 136}, 183 (1984).
  doi:10.1016/0370-2693(84)91177-8

%\cite{Kaplan:1983sm}
\bibitem{Kaplan:1983sm} 
  D.~B.~Kaplan, H.~Georgi and S.~Dimopoulos,
  %``Composite Higgs Scalars,''
  Phys.\ Lett.\ B {\bf 136}, 187 (1984).
  doi:10.1016/0370-2693(84)91178-X

%\cite{Kaplan:1991dc}
\bibitem{Kaplan:1991dc} 
  D.~B.~Kaplan,
  %``Flavor at SSC energies: A New mechanism for dynamically generated fermion masses,''
  Nucl.\ Phys.\ B {\bf 365}, 259 (1991).
  doi:10.1016/S0550-3213(05)80021-5

%\cite{Agashe:2004rs}
\bibitem{Agashe:2004rs} 
  K.~Agashe, R.~Contino and A.~Pomarol,
  %``The Minimal composite Higgs model,''
  Nucl.\ Phys.\ B {\bf 719}, 165 (2005)
  doi:10.1016/j.nuclphysb.2005.04.035
  [hep-ph/0412089].

%\cite{Contino:2006qr}
\bibitem{Contino:2006qr} 
  R.~Contino, L.~Da Rold and A.~Pomarol,
  %``Light custodians in natural composite Higgs models,''
  Phys.\ Rev.\ D {\bf 75}, 055014 (2007)
  doi:10.1103/PhysRevD.75.055014
  [hep-ph/0612048].

%\cite{Contino:2006nn}
\bibitem{Contino:2006nn} 
  R.~Contino, T.~Kramer, M.~Son and R.~Sundrum,
  %``Warped/composite phenomenology simplified,''
  JHEP {\bf 0705}, 074 (2007)
  doi:10.1088/1126-6708/2007/05/074
  [hep-ph/0612180].

%\cite{Carena:2007ua}
\bibitem{Carena:2007ua} 
  M.~Carena, E.~Ponton, J.~Santiago and C.~E.~M.~Wagner,
  %``Electroweak constraints on warped models with custodial symmetry,''
  Phys.\ Rev.\ D {\bf 76}, 035006 (2007)
  doi:10.1103/PhysRevD.76.035006
  [hep-ph/0701055].

%\cite{Smith:1979rz}
\bibitem{Smith:1979rz} 
  P.~F.~Smith and J.~R.~J.~Bennett,
  %``A Search For Heavy Stable Particles,''
  Nucl.\ Phys.\ B {\bf 149}, 525 (1979).
  doi:10.1016/0550-3213(79)90006-3

%\cite{Smith:1982qu}
\bibitem{Smith:1982qu} 
  P.~F.~Smith, J.~R.~J.~Bennett, G.~J.~Homer, J.~D.~Lewin, H.~E.~Walford and W.~A.~Smith,
  %``A Search For Anomalous Hydrogen In Enriched D-2 O, Using A Time-of-flight Spectrometer,''
  Nucl.\ Phys.\ B {\bf 206}, 333 (1982).
  doi:10.1016/0550-3213(82)90271-1

%\cite{Hemmick:1989ns}
\bibitem{Hemmick:1989ns} 
  T.~K.~Hemmick {\it et al.},
  %``A Search for Anomalously Heavy Isotopes of Low $Z$ Nuclei,''
  Phys.\ Rev.\ D {\bf 41}, 2074 (1990).
  doi:10.1103/PhysRevD.41.2074

%\cite{Starkman:1990nj}
\bibitem{Starkman:1990nj} 
  G.~D.~Starkman, A.~Gould, R.~Esmailzadeh and S.~Dimopoulos,
  %``Opening the Window on Strongly Interacting Dark Matter,''
  Phys.\ Rev.\ D {\bf 41}, 3594 (1990).
  doi:10.1103/PhysRevD.41.3594

%\cite{Verkerk:1991jf}
\bibitem{Verkerk:1991jf} 
  P.~Verkerk, G.~Grynberg, B.~Pichard, M.~Spiro, S.~Zylberajch, M.~E.~Goldberg and P.~Fayet,
  %``Search for superheavy hydrogen in sea water,''
  Phys.\ Rev.\ Lett.\  {\bf 68}, 1116 (1992).
  doi:10.1103/PhysRevLett.68.1116

%\cite{Yamagata:1993jq}
\bibitem{Yamagata:1993jq} 
  T.~Yamagata, Y.~Takamori and H.~Utsunomiya,
  %``Search for anomalously heavy hydrogen in deep sea water at 4000-m,''
  Phys.\ Rev.\ D {\bf 47}, 1231 (1993).
  doi:10.1103/PhysRevD.47.1231

%\cite{Kudo:2001ie}
\bibitem{Kudo:2001ie} 
  A.~Kudo and M.~Yamaguchi,
  %``Inflation with low reheat temperature and cosmological constraint on stable charged massive particles,''
  Phys.\ Lett.\ B {\bf 516}, 151 (2001)
  doi:10.1016/S0370-2693(01)00938-8
  [hep-ph/0103272].

%\cite{Mack:2007xj}
\bibitem{Mack:2007xj} 
  G.~D.~Mack, J.~F.~Beacom and G.~Bertone,
  %``Towards Closing the Window on Strongly Interacting Dark Matter: Far-Reaching Constraints from Earth's Heat Flow,''
  Phys.\ Rev.\ D {\bf 76}, 043523 (2007)
  doi:10.1103/PhysRevD.76.043523
  [arXiv:0705.4298 [astro-ph]].

%\cite{Mack:2012ju}
\bibitem{Mack:2012ju} 
  G.~D.~Mack and A.~Manohar,
  %``Closing the window on high-mass strongly interacting dark matter,''
  J.\ Phys.\ G {\bf 40}, 115202 (2013)
  doi:10.1088/0954-3899/40/11/115202
  [arXiv:1211.1951 [astro-ph.CO]].

\bibitem{Aad:2014efa} 
  G.~Aad {\it et al.} [ATLAS Collaboration],
  %``Search for pair and single production of new heavy quarks that
	%decay to a $Z$ boson and a third-generation quark in $pp$
	%collisions at $\sqrt{s}=8$ TeV with the ATLAS detector,''
  JHEP {\bf 1411}, 104 (2014)
 doi:10.1007/JHEP11(2014)104
  [arXiv:1409.5500 [hep-ex]].

\bibitem{Aad:2015gdg} 
  G.~Aad {\it et al.} [ATLAS Collaboration],
  %``Analysis of events with $b$-jets and a pair of leptons of the same
	%charge in $pp$ collisions at $\sqrt{s}=8$ TeV with the ATLAS
	%detector,''
  JHEP {\bf 1510}, 150 (2015)
  doi:10.1007/JHEP10(2015)150
  [arXiv:1504.04605 [hep-ex]].

\bibitem{Aad:2015kqa} 
  G.~Aad {\it et al.} [ATLAS Collaboration],
  %``Search for production of vector-like quark pairs and of four top
	%quarks in the lepton-plus-jets final state in $pp$ collisions
	%at $\sqrt{s}=8$ TeV with the ATLAS detector,''
  JHEP {\bf 1508}, 105 (2015)
  doi:10.1007/JHEP08(2015)105
  [arXiv:1505.04306 [hep-ex]].

\bibitem{Buchkremer:2013bha} 
  M.~Buchkremer, G.~Cacciapaglia, A.~Deandrea and L.~Panizzi,
  %``Model Independent Framework for Searches of Top Partners,''
  Nucl.\ Phys.\ B {\bf 876}, 376 (2013)
  doi:10.1016/j.nuclphysb.2013.08.010
  [arXiv:1305.4172 [hep-ph]].

%\cite{Fajfer:2013wca}
\bibitem{Fajfer:2013wca} 
  S.~Fajfer, A.~Greljo, J.~F.~Kamenik and I.~Mustac,
  %``Light Higgs and Vector-like Quarks without Prejudice,''
  JHEP {\bf 1307}, 155 (2013)
  doi:10.1007/JHEP07(2013)155
  [arXiv:1304.4219 [hep-ph]].

\bibitem{Alwall:2014hca} 
  J.~Alwall {\it et al.},
  %``The automated computation of tree-level and next-to-leading order
	%differential cross sections, and their matching to parton
	%shower simulations,''
  JHEP {\bf 1407}, 079 (2014)
  doi:10.1007/JHEP07(2014)079
  [arXiv:1405.0301 [hep-ph]].

\bibitem{VLQmodel}
https://feynrules.irmp.ucl.ac.be/wiki/VLQ

\bibitem{Alloul:2013bka} 
  A.~Alloul, N.~D.~Christensen, C.~Degrande, C.~Duhr and B.~Fuks,
  %``FeynRules  2.0 - A complete toolbox for tree-level phenomenology,''
  Comput.\ Phys.\ Commun.\  {\bf 185}, 2250 (2014)
  doi:10.1016/j.cpc.2014.04.012
  [arXiv:1310.1921 [hep-ph]].

\bibitem{Sjostrand:2006za} 
  T.~Sjostrand, S.~Mrenna and P.~Z.~Skands,
  %``PYTHIA 6.4 Physics and Manual,''
  JHEP {\bf 0605}, 026 (2006)
  doi:10.1088/1126-6708/2006/05/026
  [arXiv:hep-ph/0603175].

\bibitem{Drees:2013wra} 
  M.~Drees, H.~Dreiner, D.~Schmeier, J.~Tattersall and J.~S.~Kim,
  %``CheckMATE: Confronting your Favourite New Physics Model with LHC
	%Data,''
  Comput.\ Phys.\ Commun.\  {\bf 187}, 227 (2014)
  doi:10.1016/j.cpc.2014.10.018
  [arXiv:1312.2591 [hep-ph]].

\bibitem{deFavereau:2013fsa} 
  J.~de Favereau {\it et al.} [DELPHES 3 Collaboration],
  %``DELPHES 3, A modular framework for fast simulation of a generic
	%collider experiment,''
  JHEP {\bf 1402}, 057 (2014)
  doi:10.1007/JHEP02(2014)057
  [arXiv:1307.6346 [hep-ex]].

\bibitem{Aliev:2010zk} 
  M.~Aliev, H.~Lacker, U.~Langenfeld, S.~Moch, P.~Uwer and
	M.~Wiedermann,
  %``HATHOR: HAdronic Top and Heavy quarks crOss section calculatoR,''
  Comput.\ Phys.\ Commun.\  {\bf 182}, 1034 (2011)
  doi:10.1016/j.cpc.2010.12.040
  [arXiv:1007.1327 [hep-ph]].

%\cite{Martin:2009iq}
\bibitem{Martin:2009iq} 
  A.~D.~Martin, W.~J.~Stirling, R.~S.~Thorne and G.~Watt,
  %``Parton distributions for the LHC,''
  Eur.\ Phys.\ J.\ C {\bf 63}, 189 (2009)
  doi:10.1140/epjc/s10052-009-1072-5
  [arXiv:0901.0002 [hep-ph]].

\bibitem{Aad:2014wea} 
  G.~Aad {\it et al.} [ATLAS Collaboration],
  %``Search for squarks and gluinos with the ATLAS detector in final
	%states with jets and missing transverse momentum using
	%$\sqrt{s}=8$ TeV proton--proton collision data,''
  JHEP {\bf 1409}, 176 (2014)
  doi:10.1007/JHEP09(2014)176
  [arXiv:1405.7875 [hep-ex]].

\end{thebibliography}
\end{document}